\documentclass[aps,prd,reprint,groupedaddress]{revtex4-2}
\input epsf.sty
\usepackage{graphicx,amsmath,amssymb}

\def\lromn#1{\uppercase\expandafter{\romannumeral#1}}

\begin{document}


\title{
Dynamical relaxation of cosmological constant 
}



\author{M. Yoshimura}
\affiliation{Research Institute for Interdisciplinary Science,
Okayama University \\
Tsushima-naka 3-1-1 Kita-ku Okayama
700-8530 Japan}


\date{\today}

\begin{abstract}

A special class of conformal gravity theories is proposed to
solve the long standing problem of the fine-tuned cosmological constant. 
In the proposed model  time evolution of the inflaton field
leaves behind a nearly vanishing, but finite value of dark energy density
of order (a few meV)$^4$ to explain the late-time accelerating universe.
A multiple scalar inflaton field is assumed 
to have a conformal coupling to the Ricci scalar
curvature in the lagrantian, which results in, after a Weyl rescaling
to the Einstein metric frame, modification of
inflaton kinetic and potential terms along with its coupling to 
Higgs fields in the standard model.
One may define an effective cosmological $\Lambda$
function in the Einstein metric frame,
 which controls, when the potential is added,
 a slow-roll inflation and subsequent oscillation 
at the potential minimum of a spontaneous symmetry breaking phase.
The inflaton oscillation accompanies 
 particle production towards thermalized hot big-bang universe.
At the same time zero-point quantum fluctuation
of second inflaton field is generated,  and its accumulated fluctuation gives rise to a
symmetry restoration, pushing back the inflaton field towards the infinity.
This gives a dynamical relaxation  of 
vanishing effective cosmological constant.
Both inhomogeneous inflaton modes and their collapsed black holes
of primordial origin are good candidates of cold dark matter.

\end{abstract}


\maketitle

\section{Introduction}

There exist a vast literature  of proposals,
\cite{lambda-problem}, \cite{k-essence}, \cite{bousso-polchinski}, 
\cite{witten}, \cite{cc polyakov},  to resolve
the fine-tuned cosmological constant problem.
We would like to add a new idea to this list.
If our idea is correct, it requires a flesh view
of the early universe, as explained in the present and subsequent works.

We need to combine a few concepts in order to construct our framework,
hence we first explain how they come about.

(1) In order to regard the cosmological constant as dynamical, we introduce
conformal coupling to the Ricci scalar curvature $R$ in the form,
$ - M_{\rm P}^2 F(\chi) R$, in the lagrangian density.  Here
$\chi$ is the inflaton field and $M_{\rm P} \approx 10^{18}$GeV is the Planck energy.
After the Weyl rescaling of the metric tensor to eliminate $F(\chi)$,
 the cosmological constant $\Lambda$ is changed to a functional variable
$\Lambda_{\rm eff} =\Lambda /F^2(\chi)$.
With a choice of appropriate polynomial for $F(\chi)$, 
this dynamical cosmological constant becomes null at $\chi \rightarrow \infty$.

(2) Slow-roll inflation is realized giving a right amount of spectral index and
tensor-to-scalar mode ratio consistent with observations.
Inflation ends with Higgs boson production followed by
quick thermalization among standard model particles by their coupling to
Higgs boson.
The particle production is described as a parametric amplification of time
dependent harmonic oscillator system.
Besides  interaction term of the form, $ \chi^2 H^2$ to the Higgs boson $H$,
the inflaton $\chi$ also couples to another scalar field $\phi$.
$\chi$ field oscillation occurs around a potential minimum of $V(\chi)$ away from
the origin at $\chi=0$ in which Nambu-Goldstone kinetic repulsion
dominates \cite{my21}. It is driven by a Higgs-like potential 
$V(\chi) = g (\chi^2 +\phi^2 - v^2)^2/4\,, g > 0$.
$\phi$ production is necessarily
 accompanied by quantum zero-point fluctuation $\langle \phi^2 \rangle$.
This induces a positive contribution to $\chi$ mass  like
$g \langle \phi^2 \rangle$, adding 
to the original negative term $- g v^2 \equiv - m_{\chi}^2 < 0$.

(3) The leading behavior of the effective potential contains 
a piece $( - m_{\chi}^2 + g \langle \phi^2 \rangle) /\chi^2$.
Hence the potential minimum disappears at 
a critical point, $ g \langle \phi^2 \rangle = m_{\chi}^2$, 
signaling a kind of phase transition.
Some time after this critical fluctuation is reached, the $\chi$ field leave
the potential minimum, and  rolls to the field infinity governed by $- \ln \chi$ potential,
giving essentially the vanishing cosmological constant. 

(4) Spatially homogeneous part constitutes dark energy whose energy
density at present becomes of order $(H_0 M_{\rm P})^2 \sim ({\rm a\; few\; meV})^4$.
At the same time
field quanta of inhomogeneous modes are thermally
produced from ambient medium, and shortly after the electroweak phase transition
they behave as non-relativistic cold dark matter (CDM).
Gravitationally collapsed clumps of inhomogeneous modes
are another good candidate of cold dark matter.

The rest of this paper is organized as shown in the following table of contents.

=============================

\vspace{0.3cm}
\lromn2. 
Conformal coupling to gravity and dynamical cosmological ``constant''

A. 
Conformal gravity

B.  Choice of conformal coupling function  and potential 

C.  Spontaneous symmetry breaking and Nambu-Goldstone kinetic modes

D. Inflaton coupling to Higgs boson

\vspace{0.3cm}
\lromn3.
Time evolution at inflationary epoch

A.
Field and Einstein equations

B.
Slow-roll inflation

C.
Spectral index and tensor-to-scalar ratio

D.
Oscillation at the end of inflation

\vspace{0.3cm}
\lromn4.
Particle production and symmetry restoration

A. Particle production and quantum fluctuation due to inflaton oscillation

B. Radiation dominated epoch

\vspace{0.3cm}
\lromn5.
Late time acceleration

A.
Differential equations and its solutions

B.
Equation-of-state factor

\vspace{0.3cm}
\lromn6.
Inhomogeneous inflaton modes and candidate cold dark matter

A.
Inhomogeneous inflaton modes

B.
Time variation of CDM energy density

\vspace{0.3cm}
\lromn7.
Summary

\vspace{0.3cm}
\lromn8.
Appendix: Production rate and decay freeze-out of inhomogeneous inflaton modes

=============================
\vspace{0.3cm}
We use the natural unit of $\hbar = c  = 1$ and the Boltzmann constant $ k_B = 1$
throughout the present work  unless otherwise stated.
The cosmic scale factor $a(t)$ is introduced in
the flat Friedman-Robertson-Walker metric, $ds^2 = dt^2 - a^2(t) (d\vec{x})^2$
\cite{cosmology}.

\section
{\bf Conformal coupling to gravity and dynamical cosmological ``constant''}

\subsection
{\bf Conformal gravity}

We would like to treat the cosmological constant $\Lambda$ as
a  dynamical variable $\Lambda_{\rm eff}(t)$.
Let us explore for this purpose conformal coupling to gravity
in which $\Lambda$ term is multiplied by a function of inflaton field
after a Weyl rescaling.

The simplest conformal gravity scheme  uses a single real inflaton field $\chi$,
and the next simplest scheme for our purpose uses two complex fields, the inflaton
and an additional scalar field $\phi$.
In this subsection we shall deal with the first case.
In the Jordan metric frame 
\cite{jbd} the lagrangian density has a generic form,
\begin{eqnarray}
&&
{\cal L} = {\cal L}_{\chi g}^{(J)} + {\cal L}_{\rm SM} 
\,, \hspace{0.5cm}
{\cal L}_{\chi g}^{(J)} =  \sqrt{-g}  \times
\nonumber \\ &&
\left(  - M_{\rm P}^2 F(\chi) R(g_{\mu \nu})
- 2 M_{\rm P}^2\, \Lambda
+ \frac{1}{2} (\partial \chi)^2 - V(\chi)
\right)
\,,
\label {lagrangian in Jordan frame}
\\ &&
{\cal L}_{\rm SM} =  \sqrt{-g} L_{\rm SM} (\psi, g_{\mu \nu})
\,,
\\ &&
M_{\rm P}^2 = \frac{1}{16 \pi G_N} \sim (1.72 \times 10^{18}\, {\rm GeV})^2
\,, \hspace{0.5cm}
\Lambda > 0
\,.
\end{eqnarray}
$L_{\rm SM}(\psi, g_{\mu \nu}) $ is the standard model lagrangian density with $\psi$
fields of standard theory; gauge bosons, Higgs boson, and fermions (quarks and leptons),  
generically.
There are two functional degrees of freedom;
 $F(\chi)$ and $V(\chi)$ in the scalar-tensor gravity
given by ${\cal L}_{\chi g}^{(J)}  $.
These are called the conformal
couping function and the potential function.
We restrict these to polynomial forms, and  assume
a reflection symmetry $F(-\chi) = F(\chi)$ and $V(-\chi) = V(\chi)$,
hence they are even polynomial functions of $\chi^2$.
The Jordan frame is useful if four dimensional conformal gravity
descends from higher dimensional theories such as
Kaluza-Klein unification \cite{kaluza-klein}, \cite{gh unification}
and superstring theories.

It is physically more transparent in cosmological applications
to transform the
lagrangian density in the Jordan frame to that in  the Einstein metric frame
by a Weyl-rescaling, $\bar{g}_{\mu \nu} = F(\chi) g_{\mu \nu}$.
To simplify our notation, we replace  the new metric $\bar{g}_{\mu \nu}  $ 
by $g_{\mu\nu}$.
The lagrangian density in the new Einstein frame is then
\begin{eqnarray}
&&
{\cal L}_{\chi g}^{(E)} = \sqrt{-g} \left(  - M_{\rm P}^2 R(g_{\mu \nu})
+ \frac{5}{2 F(\chi)} (\partial \chi)^2 
\right.
\nonumber \\ &&
\left.
- \frac{1}{F^2(\chi)} 
(2 M_{\rm P}^2\, \Lambda + V(\chi) \right)
\,.
\label {e-frame chi-g}
\end{eqnarray}
The metric is transformed as well according to
$ds^2 = F^{-1} (dt^2 - a^2 (d\vec{x})^2 ) $.
One can re-define field variable to $\bar{\chi}$, though not necessary,
 to obtain the standard, normalized form of kinetic term, $ 5 (\partial \chi)^2/(2 F) 
= (\partial \bar{\chi})^2/2$, leaving a complicated potential function.
But we keep the original form above.

We may define a dynamical cosmological variable 
$\Lambda_{\rm eff}(\chi)$ by
\begin{eqnarray}
&&
V_{\chi g}^{(E)} = 2 M_{\rm P}^2 \Lambda_{\rm eff}(\chi) +
 \frac{V(\chi)}{ F^2(\chi)}
\,,
\\ &&
\Lambda_{\rm eff}(\chi) = \frac{\Lambda}{ F^2(\chi)}
\,.
\end{eqnarray}
which appears in (\ref{e-frame chi-g}).
It should be clear that $ \Lambda_{\rm eff}(\chi) $,
and not $ \Lambda$, should vanish when the problem
of cosmological constant is solved.
It is found later that  cosmological evolution  drives
$\chi$ towards the infinity.
It is then clear that  the maximum power of polynomial ratio $ V(\chi)/F^2(\chi)$ 
becomes important.
Suppose that the maximum powers of the two functions, $V\,, F$ are
$2n$ and $2l$ such that $ V(\chi)/F^2(\chi) \rightarrow \chi^{2n - 4l }$.
It is found that $l = n$ case gives a positive force term $\propto 1/\chi$
to the inflaton field equation,
and a dynamical cosmological constant $\propto 1/\chi^{2l}$.
We shall work out most extensively
 the case of $l=n=2$, quartic conformal and potential functions
denoted by $(F_4, V_4)$.

\subsection
{\bf Choice of conformal coupling function  and potential}

We parametrize the conformal coupling function and the potential  as
\begin{eqnarray}
&&
V = V_4(\chi) = - \frac{m_{\chi}^2}{2} \chi^2 + \frac{g}{4} \chi^4
\,, 
\\ &&
F =
F_4 (\chi) = 1 + \xi_2 (\frac{\chi}{M_{\rm P}})^2 + \xi_4 (\frac{\chi}{M_{\rm P}})^4
\,. 
\end{eqnarray}
We arranged parameters of $m_{\chi}^2 > 0$ such that a spontaneous breaking of
a discrete symmetry occurs at $\chi^2 = m_{\chi}^2/g$.
All other constant parameters are taken positive.

It would be useful to record  limiting behaviors of the effective cosmological constant
$\Lambda_{\rm eff}(\chi) $
in the $  (F_4, V_4)$ model, at small  and large fields:
\begin{eqnarray}
&&
V_{\chi g}^{(E)} 
\rightarrow \Lambda 
- \frac{m_{\chi}^2 + 8 \xi_2 \Lambda}{4 M_{\rm P}^2} 
\chi^2 + O( \chi^4)
 \,, \hspace{0.3cm}
{\rm as} \; \chi \rightarrow 0
\,,
\\ &&
V_{\chi g}^{(E)}  \rightarrow \frac{g}{8 \xi_4^2} \frac{ M_{\rm P}^6 }{\chi^4}
+ O(\chi^{-6})
\,, \hspace{0.3cm}
{\rm as}\; \chi \rightarrow \infty
\,.
\end{eqnarray}
Note that the original cosmological constant disappears in
the infinite field limit.

\subsection
{\bf Spontaneous symmetry breaking and Nambu-Goldstone kinetic modes
}

So far we considered a discrete reflection symmetry
of a real inflaton field.
What is  important in resolution of the cosmological constant problem
is, however, a spontaneously broken continuous symmetry, which
gives rise to Nambu-Goldstone modes.
But if the inflaton field after inflation
 is permanently trapped in the potential minimum implied by the broken phase,
there may exist an unacceptably large cosmological constant.
To avoid this catastrophe, 
the potential minimum must change towards the vanishing effective
cosmological constant at the field infinity.
This is made possible if a higher continuous symmetry
exists and a quantum fluctuation of another inflaton field develops
such that  symmetry restoration becomes possible.
The simplest realization of this scenario is to choose
 O(4) symmetry breaking.

We shall start from the simplest O(2) symmetric model
from a pedagogical reason.
In two-component field space of $(\chi_1, \chi_2)$
an abstract angular momentum may be defined by
$\chi^2 \dot{\theta} $ where $ (\chi_1, \chi_2) = \chi (\cos \theta\,, \sin \theta)$.
The angular field equation of this inflaton system gives
$\partial_t ( a^3 \chi^2 \dot{\theta} ) =0 $.
This equation has a trivial solution;
$\chi^2 \dot{\theta} = c_{\chi}/a^3$ with $c_{\chi}$ a constant of integration.
This solution may be incorporated as a centrifugal repulsion term
in an effective potential, $c_{\chi}^2/(2 a^6 \chi^2)$ \cite{my21}.

In O(3) extension the invariant angular momentum operator
is uniquely defined, and this single operator is not sufficient
to ensure symmetry restoration for resolution of the cosmological
constant problem.
The next extension to O(4) symmetry introduces two invariant angular momenta,
since this group is locally isomorphic to O$_+$(3) $\times $ O$_-$(3).
The generators of this maximal subgroup are 
$(L_{41} \pm L_{23})/2\,, (L_{42} \pm L_{31})/2\,,  (L_{43} \pm L_{12})/2$.
Using four real inflaton fields, $\chi_i\,, i = 1 \sim 4$,
one finds that two invariant squared fields denoted by $\chi^2,\phi^2$ exist:
\begin{eqnarray}
&&
(\chi^2\,, \phi^2) =
3 \sum_{i=1}^4 \chi_i^2 \pm 2 (\chi_1 \chi_2 + \chi_2 \chi_3 + \chi_3 \chi_1)
\,.
\end{eqnarray}
We shall assume $\chi^2 > \phi^2$, calling them large and small components.
Using these, one derives the centrifugal repulsive potential in O(4) symmetric
model:
\begin{eqnarray}
&&
\frac{c_+^2} { 2 a^6 \chi^2} + \frac{c_-^2} { 2 a^6 \phi^2}
\,.
\end{eqnarray}

Our new potential for two inflaton fields is, in the Jordan frame,
\begin{eqnarray}
&&
\hspace*{1cm}
V^{(J)}(\chi, \phi) = 
\nonumber \\ &&
\frac{c_+^2} { 2 a^6 \chi^2} + \frac{c_-^2} { 2 a^6 \phi^2}
+ \frac{g}{4} (\chi^2 + \phi^2 - \frac{m_{\chi}^2}{g} )^2 + 2 M_{\rm P}^2 \Lambda 
\,.
\label {o4 centrifugal repulsion}
\end{eqnarray}
The centrifugal repulsion acts as a potential wall at field zero, 
and its effect becomes important at
late times.

\subsection
{\bf Inflaton coupling to Higgs boson}

An appealing feature of $F-$factor  is that it automatically introduces inflaton
coupling to standard model particles, in particular to the Higgs doublet $H$:
\begin{eqnarray}
&&
V_H^{(E)} =\frac{\lambda_H}{4} \,  F^{- 2} (| H|^2 - v^2 )^2
\,.
\label {higgs-inflaton coupling}
\end{eqnarray}
With extended two $(\chi, \phi)$ real fields 
of O(4) scheme a
reduced O(2) symmetric field combination  $\bar{\chi}^2 = \chi^2+\phi^2$ 
 appears in the variable function, $ F(\bar{\chi}^2)$.
Both Higgs boson mass and inflaton coupling to Higgs boson are determined by
eq.(\ref{higgs-inflaton coupling}), and they are dynamical, meaning that
these quantities are dependent on cosmic times, or the redshift factor
$z+1 = a(t_0)/a(t)$.

We need to separate particle quantum field parts, $\chi\,, \phi$, 
from the c-number background field denoted by $\chi_b(t)$.
This separation leads to
\begin{eqnarray}
&&
\bar{\chi}^2 = \chi_b^2 + 2 \chi_b \chi + \chi^2 + \phi^2
\,, \hspace{0.3cm}
\chi_b \gg \sqrt{\chi^2 + \phi^2 }
\,,
\\ &&
F^{- 2} = \left( \xi_4 \frac{\chi_b^4}{M_{\rm P}^4}\right)^{- 2} \left(
1 - 8  \frac{\chi}{\chi_b} + O ( \frac{\chi^2}{\chi_b^2}\,, \frac{\phi^2}{\chi_b^2} )
\right)
\,.
\end{eqnarray}
The physical Higgs field $H_0$ is identified after the electroweak phase transition by
$H = (H_0 + v\,, 0)$, the rest of doublet components being absorbed as
longitudinal parts of electroweak gauge bosons.
Hence, when the separation of inflaton field is applied to (\ref{higgs-inflaton coupling}),
the  potential relevant to the Higgs boson mass is
\begin{eqnarray}
&&
m_H = 125\, {\rm GeV}\, \left(\xi_4 (\frac{\chi_b}{M_{\rm P}})^4  \right)^{- 1}
\,.
\label {higgs boson mass vs chi}
\end{eqnarray}
The large field limit formula was used here. 
Note that the last factor in (\ref{higgs boson mass vs chi}) is caused by 
the metric change due to $F$ entailing
the spacetime coordinate change.
This further modifies the unit change of energy and mass, hence
it is not really a change of physical mass.

Three-point vertexes of $\chi \phi \phi\,, \chi H_0 H_0$ 
are derived from the same equation (\ref{higgs-inflaton coupling}); 
\begin{eqnarray}
&&
V_{\chi \phi H}^{(E)} =  c_{\phi} \chi \phi^2 + c_{H} \chi H_0^2
\,, 
\\ &&
c_{\phi}  = g \chi_b
\,, \hspace{0.3cm}
c_{H} =  - 8  \lambda_H \frac{v^2}{\chi_b} (\xi_4
\frac{\chi_b^4}{M_{\rm P}^4} )^{-2} 
\,.
\label {chi-coupling}
\end{eqnarray}
This lagrangian density can describe $\chi$ production processes
via virtual Higgs-exchange diagram,
$l l \rightarrow l l \chi \,, q q \rightarrow q q \chi$ with
$l\,, q$ being leptons and quarks,
and $\chi$ decay process,
$\chi \rightarrow \phi \phi \,, \chi \rightarrow H_0 H_0\,, H_0 l \bar{l}\,, H_0 q \bar{q}$,
depending on the mutual mass relation between $\chi$ and $H_0$.
Redshift dependence of  background $\chi_b-$field shall be calculated
to give production and decay rates in subsequent sections.

\section
{\bf Time evolution at inflationary epoch}

\subsection
{\bf Field and Einstein equations}

For discussion of inflationary epoch only $\chi$ field is important,
and we shall retain relevant parts alone.
Thus, inflaton coupling to the Higgs field can be neglected.

The general form of the inflaton field equation is
derived using the variational principle applied to ${\cal L}_{\chi g}^{(E)}$
of eq.(\ref{e-frame chi-g}):
\begin{eqnarray}
&&
5 \left( \partial_t \frac{\partial_t \chi}{F(\chi)}  + 3 \frac{\dot{a}}{a} 
\frac{\partial_t \chi}{F(\chi) }  
-  \frac{1}{a^2} \vec{\nabla} \frac{ \vec{\nabla} \chi}{ F(\chi)}  \right)
= - \frac{\partial }{\partial \chi} V_{\chi g}^{(E)} (\chi)
\,.
\label {chi-field eq}
\nonumber \\ &&
\end{eqnarray}

We shall first discuss the spatially homogeneous mode, assuming
$\chi$ field depending on time alone.
Using time derivative $d (1/F)/dt  = - \dot{\chi} \partial_{\chi} F/F^2$,
the field  equation is
\begin{eqnarray}
&&
\ddot{\chi} +  3 \frac{\dot{a}}{a} \dot{\chi}
-  \frac{\partial_{\chi} F}{F } \dot{\chi}^2 
= - \frac{1}{5} F \, \frac{\partial }{\partial \chi} V_{\chi g}^{(E)} (\chi)
\,,
\label {conformal chi-eq}
\\ &&
\hspace*{-0.3cm}
  F \, \frac{\partial }{\partial \chi} V_{\chi g}^{(E)} (\chi)
=
\frac{1}{ F}
\left( \partial_{\chi}V - 2 (V + 2 M_{\rm P}^2 \Lambda) \frac{\partial_{\chi} F}{F}  \right)
\,.
\end{eqnarray}
In $(F_4, V_4)$ model, the leading 
asymptotic behavior of the right-hand side of field equation 
(\ref{conformal chi-eq}) is 
\begin{eqnarray}
&&
\frac{ g}{5 \xi_4} \frac{M_{\rm P}^4}{\chi}
+ O(\chi^{-3})
\,.
\end{eqnarray}

\begin{figure*}[htbp]
 \begin{center}
 \epsfxsize=0.6\textwidth
 \centerline{\epsfbox{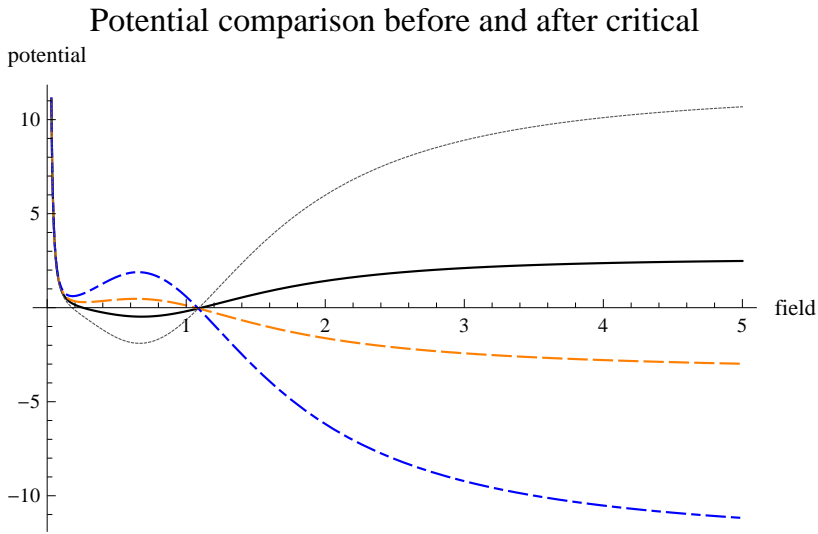}} \hspace*{\fill}\vspace*{1cm}
   \caption{
Effective potentials adding a centrifugal repulsion $0.01/\chi^2$ 
after and before critical point of quantum fluctuation in $(F_4, V_4)$ model.
Common parameters are $(M_{\rm P}, \Lambda, \xi_2,\xi_4, g ) =
(1,0.1, 1,1,1 ) $ in the Planck unit:
effective potential before critical turning point
 of $\sqrt{m_{\chi}^2 -  g \langle \phi^2 \rangle} = 5$ in solid black,
of $\sqrt{ m_{\chi}^2 -  g \langle \phi^2 \rangle} = 10$ in dotted black,
and after critical turning point of  $\sqrt{ g \langle \phi^2 \rangle - m_{\chi}^2} = 5$    in dashed orange, and 10  in dash-dotted blue.
All potentials  decreases in proportion to a negative coefficient $\times \ln \chi$
at field infinity, but existence of potential minimum depends on the next leading
term $\propto 1/\chi^2$, with coefficients changing with quantum fluctuation.
}
   \label {effective-potentials before and after}
 \end{center} 
\end{figure*}

One may define an effective potential $V_{\rm eff} $ by rewriting the right-hand side
of equation for $\ddot{\chi}  $ as $
- \partial_{\chi} V_{\rm eff} $, to give
\begin{eqnarray}
&&
 V_{\rm eff}^{(E)}(\chi)
 = \frac{1}{5} \int_{\chi_*}^{\chi}  d\chi F(\chi) \partial_{\chi} V_{\chi g}^{(E)} (\chi)
\,,
\label {effective potential}
\end{eqnarray} 
where $\chi_*$ is a few times the Planck mass $M_{\rm P}$.
This potential at late times is
illustrated for $(F_4,V_4) $ models in Fig(\ref{effective-potentials before and after}).
They have a feature of having a maximum beyond which
the derived force pushes the field towards the infinity.
The Einstein equation is given, using this effective potential, by
\begin{eqnarray}
&&
( \frac{\dot{a}}{a})^2 =  \frac{1}{6 M_{\rm P}^2} \left(
\frac{5}{2 F} \dot{\chi}^2 + \frac{5}{2 F} \frac{(\vec{\nabla} \chi)^2}{a^2 }
+ V_{\rm eff}^{(E)}(\chi)
\right)
\,,
\label {einstein eq 1}
\end{eqnarray}
while the right-hand side of field equation (\ref{conformal chi-eq}) is 
given by $ - \partial_{\chi} V_{\rm eff}^{(E)}(\chi)$

The effective potential  is approximately
\begin{eqnarray}
&&
V_{\rm eff}^{(E)}(\chi) \approx \frac{M_{\rm P}^4}{5}
\left( (\frac{ \xi_2}{\xi_4 } +  \frac{3 m_{\chi}^2 }{2 g M_{\rm P}^2 }) 
(\frac{M_{\rm P}^2}{\chi_*^2} -  \frac{M_{\rm P}^2}{\chi^2}) - 
\frac{g}{\xi_4} \ln \frac{\chi}{\chi_*}
\right)
\,,
\nonumber \\ &&
\label {effective potential at infty}
\end{eqnarray}
to the leading and the next leading orders.
$\chi_*$ is of order $M_{\rm P}$ taken as
an effective starting point of inflation.
Its precise value is not important.
The asymptotic form of field equation in $(F_4, V_4)$ model  reduces to
\begin{eqnarray}
&&
\ddot{\chi} +  3 \frac{\dot{a}}{a} \dot{\chi}
-   4 \frac{\dot{\chi}^2  }{\chi } 
=  \frac{ g}{5 \xi_4} \frac{M_{\rm P}^4}{\chi}
\,.
\label {conformal chi-eq: asymptotic}
\end{eqnarray}

A closed differential equation for the inflaton field after inserting the Hubble rate $\dot{a}/a$
derived from the Einstein equation  may be useful for numerical computations,
giving the scale factor by a subsequent  integration.
These equations, (\ref{conformal chi-eq: asymptotic}) and (\ref{einstein eq 1}),
shall be used when we discuss late-time evolution.

\subsection
{\bf Slow-roll inflation}

The de Sitter spacetime characterized by
a constant cosmological constant is not eternal in scalar-tensor gravity,
and we need to identify the epoch that realizes an approximate de Sitter
spacetime.
In order to determine this epoch, it is useful to notice
a good criterion given by the slow-roll conditions, which
place bounds on the slope of potential \cite{inflation models 1}, \cite{cosmology}:
\begin{eqnarray}
&&
| \frac{\partial_{\chi} V_{\rm eff}^{(E)} }{ V_{\rm eff}^{(E)} } | \ll \frac{1}{M_{\rm P}}
\,, \hspace{0.5cm}
| \frac{\partial_{\chi}^2 V_{\rm eff}^{(E)} }{V_{\rm eff}^{(E)} } | 
\ll \frac{3}{2 M_{\rm P}^2}
\,.
\end{eqnarray}
The potential  $V_{\rm eff}^{(E)} $ is nearly a constant and one can take its value 
$M_{\rm P}^4 (\xi_2/\xi_4 + 3 m_{\chi}^2/(2 g M_{\rm P}^2) )/5 $ in (\ref{effective potential at infty}).
The slow-role conditions above give
\begin{eqnarray}
&&
\chi \gg k M_{\rm P} \; {\rm and}\; \sqrt{\frac{2 k}{3}} M_{\rm P}
\,, \hspace{0.3cm}
k = \frac{\xi_2}{\xi_4 }+ \frac{3 m_{\chi}^2}{2 g M_{\rm P}^2}
\,.
\end{eqnarray}
There is no difficulty to impose these conditions.

\subsection
{\bf Spectral index and tensor-to-scalar ratio}

\begin{figure*}[htbp]
 \begin{center}
 \epsfxsize=0.5\textwidth
 \centerline{\epsfbox{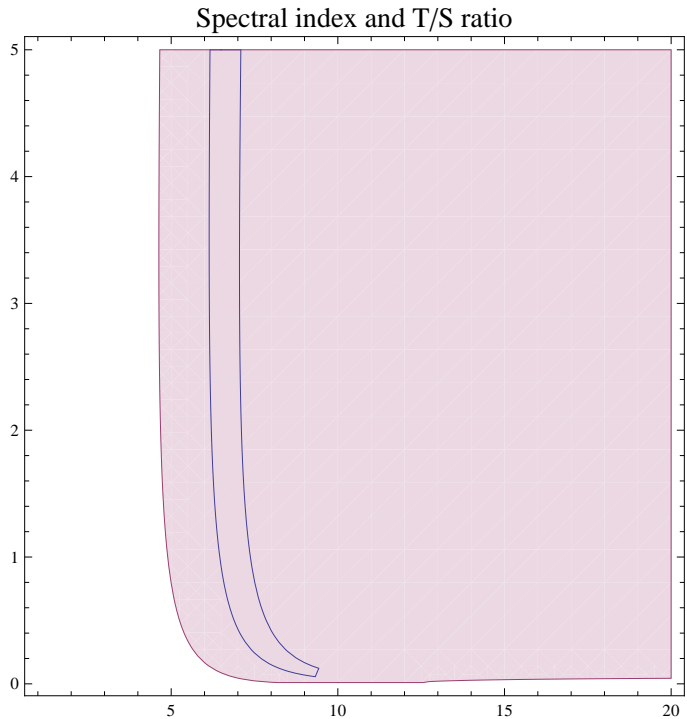}} \hspace*{\fill}\vspace*{1cm}
   \caption{
Region of spectral index in the range, $n_s -1 = 0.965 \pm 0.01$ 
(given by a thin region surrounded by two curves)
and tensor-to-scalar ratio in $r \leq 0.036$ plotted 
in $(\chi/M_{\rm P}\,, \xi_4)$ plane.
Other parameters are fixed at $\xi_2 = 1\,, g =2 \,, M_{\rm P} = 1 \,, |m_{\chi}|= 7\,,
\Lambda = 0.1 M_{\rm P}^2\,, c_{+}=5 $.
}
   \label {spectral index and ts 1}
 \end{center} 
\end{figure*}

Potential slopes actually give more than this.
When  inflaton exits horizon and re-enters
after inflation, inflaton
quantum fluctuations give seeds for the structure formation at late universe.
Two important quantities arising from this consideration are
the spectral index of density perturbation $n_s$ and the ratio
of tensor-to-scalar modes $r$.
These quantities are given in terms of potential slope parameters \cite{kami-kov}.
\begin{eqnarray}
&&
n_s = 1+ 2 M_{\rm P}^2 \left( 2 \frac{V_{\rm eff}''}{ V_{\rm eff}}  - 3 
( \frac{V_{\rm eff}'}{ V_{\rm eff} }  )^2
\right)
\,, 
\\ &&
r = 16 M_{\rm P}^2 ( \frac{V_{\rm eff}'}{V_{\rm eff} }  )^2
\,.
\end{eqnarray}
These simple formulas are valid for single inflaton model
defined by a single potential.
Our model during inflationary epoch is well approximated by
a  potential of a single inflaton field, and gives
\begin{eqnarray}
&&
\hspace*{-0.3cm}
\frac{V_{\rm eff}'}{ V_{\rm eff} } \approx - \frac{g}{ k \xi_4 }\frac{1}{\xi_*}
\,, \hspace{0.3cm}
\frac{V_{\rm eff}''}{ V_{\rm eff}} \approx \frac{g}{ k \xi_4 }
\left( \frac{1}{\xi_*^2} - \frac{  6 k \xi_4}{ g} \frac{M_{\rm P}^2 }{\xi_*^4 } \right)
\,.
\end{eqnarray}
To leading large field approximation, the spectral index and the tensor-to-scalar ratio are
related by
\begin{eqnarray}
&&
n_s - 1 + \frac{3}{8} r \approx
4 M_{\rm P}^2  \frac{V_{\rm eff}''}{ V_{\rm eff}} 
\,.
\end{eqnarray}
Observations have given us $n_s \sim 0.965\pm 0.01$ and $r \leq 0.036$
\cite{planck}, \cite{bicep/keck}.
Our predictions for $n_s-1\,, r$
computed using accurate potential and derivative functions
are illustrated in Fig(\ref{spectral index and ts 1}).
Similar plots in other two-parameter planes may also be depicted.
There are good chances that observed values are well reproduced in our model
if initial $\chi$ values at inflation
are within a slow-roll range of $(7\sim 8) M_{\rm P}$.

\subsection
{\bf Oscillation at the end of inflation}

The presence of potential minimum in $V_{\rm eff}^{(E)}(\chi)$ with spontaneously broken
 Nambu-Goldstone centrifugal force provides a shifted potential
minimum away from the origin. 
The field equation has an approximate form in $(F_4, V_4)$ model,
\begin{eqnarray}
&&
\ddot{\chi} + 3 \frac{\dot{a}}{a} \dot{\chi} - 4 \frac{\dot{\chi}^2}{\chi} = 
- \partial_{\chi} V_{\rm eff}^{(E)} 
\,,
\\ &&
- \partial_{\chi} V_{\rm eff}^{(E)}  
\approx  M_{\rm P}^4 \left(
\frac{c_1}{ \chi} -  \frac{c_3}{ \chi^3} 
\right)
\,, 
\\ &&
c_1 = \frac{g   }{5 \xi_4 }
\,, \hspace{0.3cm}
c_{3} = \frac{ 3 \xi_4 m_{\chi}^2+ 2 g \xi_2  M_{\rm P}^2 }{5 \xi_4^2}
\,.
\label {potential term coeffs}
\end{eqnarray}
The potential form with $m_{\chi} \neq 0$ 
is qualitatively the same as for the massless $\chi$ field
of purely quartic coupling, namely the case of $m_{\chi}=0$.
The precise value of parameter $m_{\chi}$ is not important in $c_3$.

Oscillation period is characterized by the minimum  field
point, $\chi_b = \chi_e$(a constant)  and oscillation frequency $\omega$,
\begin{eqnarray}
&&
\chi_e = \sqrt{\frac{c_{3}}{c_1}}
\,, \hspace{0.5cm}
\omega^2 = 4 M_{\rm P}^4 \frac{c_1^2}{c_3}
\,.
\end{eqnarray}
During this phase the anti-friction is 
\begin{eqnarray}
&&
\frac{\partial_{\chi} F} {F} \dot{\chi}^2 \sim 4 \frac{\dot{\chi}^2}{\chi_e}
\,,
\end{eqnarray}
giving the field equation,
\begin{eqnarray}
&&
\ddot{\chi} + 3 \frac{\dot{a}}{a}\dot{\chi} - 4 \frac{\dot{\chi}^2}{\chi_e}
\approx - \omega^2 (\chi - \chi_e)
\,.
\end{eqnarray}
If the potential curvature $\omega^2$ is large enough,
one can neglect the  anti-friction 
$\propto - \dot{\chi}^2/\chi_e$.

\section
{\bf Particle production and symmetry restoration}

Inflationary epoch ends with Higgs boson production followed by
thermalizing interaction among Higgs boson and
other standard model particles,
as outlined in \cite{preheating 2}, \cite{my95}.
We discuss in the present section another problem
related to the second inflaton.

\subsection
{\bf Particle production and quantum fluctuation due to inflaton oscillation}

We shall discuss  $\phi$ quantum production, which occurs for
a larger $\chi$ mass than a $\phi$ mass.
The quantum $\phi$ system consists of  harmonic
oscillator aggregates of its frequencies periodically oscillating 
synchronous to the $\chi$ field oscillation
around the potential minimum at $ m_{\chi}/\sqrt{g}$.
The field equation is linear in $\phi$ field and its Fourier $\vec{q}$ modes $ \phi_q$ satisfy
\begin{eqnarray}
&&
\ddot{\phi_q} + 3 \frac{\dot{a}}{a} \dot{\phi_q}  
+ \frac{q^2}{a^2} \phi_q 
= - \frac{1}{5} \frac{m_{\phi}^2 + 2 g \chi_e \chi(t)}{F} \phi_q
\,.
\end{eqnarray}
A large-amplitude oscillating $\chi(t)$ leads to parametric amplification
generally characterized as a Floquet-system, 
which produces $\phi$ quanta, as described in \cite{preheating 2}, \cite{my95}.

Particle production is necessarily accompanied by quantum fluctuation
due to fluctuation-dissipation theorem.
In \cite{my95} a formalism is developed to relate quantum fluctuation to dissipation,
namely particle production.
Their relation is given by
\begin{eqnarray}
&&
 \frac{\omega_q }{ 2} \langle \phi_q^2 \rangle =  \langle N_q \rangle \frac{d^3 q }{(2\pi)^3 }
\,.
\end{eqnarray}
Thus, the zero-point fluctuation $ \langle \phi_q^2 \rangle$ of fields is directly related to
the produced number of particles $ \langle N_q \rangle$.

\begin{figure*}[htbp]
 \begin{center}
 \centerline{\includegraphics{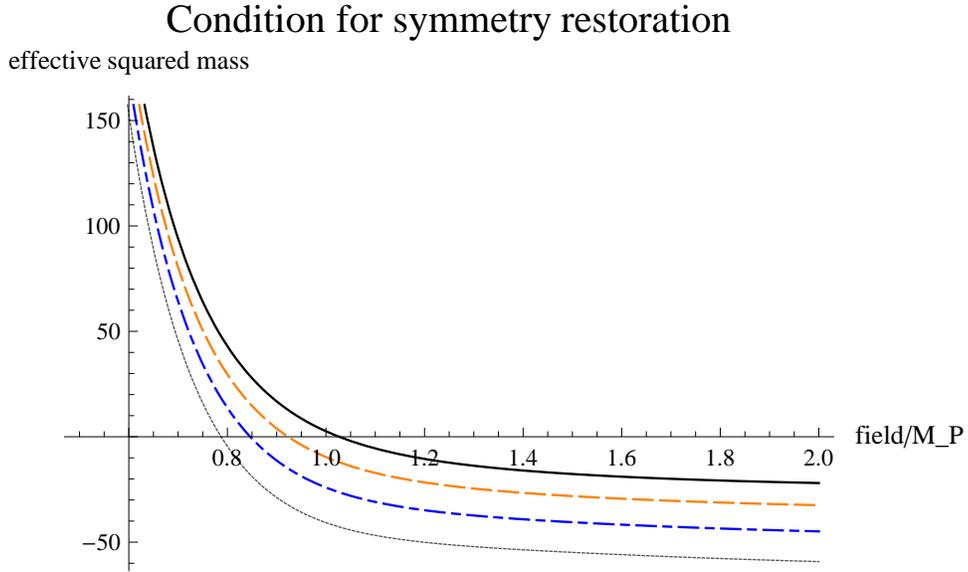}} \hspace*{\fill}
   \caption{
Left-hand side of eq.(\ref{symmetry restoration condition}) for a few choices of 
parameters: common parameters are $\xi_2=1, \xi_4=1, \Lambda = 0.1, g=5, M_{\rm P} =1, 
c_{+} =5$,
and $|m_{\chi} | = $ 5 in solid black, 6 in dashed orange, 7 in dash-dotted blue,
and 8 in dotted black.
The small field region of positive values gives symmetry restoration.
}
   \label {symmetry restored cond}
 \end{center} 
\end{figure*}

A precise condition for symmetry restoration is formulated by
calculating quantum zero-point fluctuation in terms of effective 
time-dependent $\chi$ mass $M_{\chi}$,
\begin{eqnarray}
&&
\langle \phi^2 \rangle = \frac{2}{\pi^2} \int_{m_{\phi}}^{M_{\chi}/2} d\omega_q
\sqrt{\omega_q^2 - m_{\phi}^2} \sim \frac{M_{\chi}^2}{4 \pi^2}
\,,
\\ &&
M_{\chi}^2 = - \partial_{\chi}^2 V_{\rm eff} \approx
\frac{g}{5 \xi_4} \frac{M_{\rm P}^4}{ \chi_b^2}
\,.
\label {chi-mass}
\end{eqnarray}
The condition $M_{\chi}^2 \geq 0 $ is imposed.
In this formula we assumed the massless $\phi$ for simplicity.
The condition of symmetry restoration is
\begin{eqnarray}
&&
g \langle \phi^2 \rangle - m_{\chi}^2 \geq 0
\,, \; {\rm or}\;
\frac{g}{4 \pi^2} M_{\chi}^2 - m_{\chi}^2 \geq 0
\,.
\label {symmetry restoration condition}
\end{eqnarray}
In Fig(\ref{symmetry restored cond}) we illustrate field region
in which the symmetry is restored.
The symmetry is restored to give quantum fluctuation only when
large amplitude $\chi$ oscillation is at work, namely during
copious production of $\phi$.
If accumulated amount of quantum fluctuation exceeds the
negative mass square originally set up, 
inflaton starts to roll  to the field infinity.

\begin{figure*}[htbp]
 \begin{center}
 \epsfxsize=0.6\textwidth
 \centerline{\epsfbox{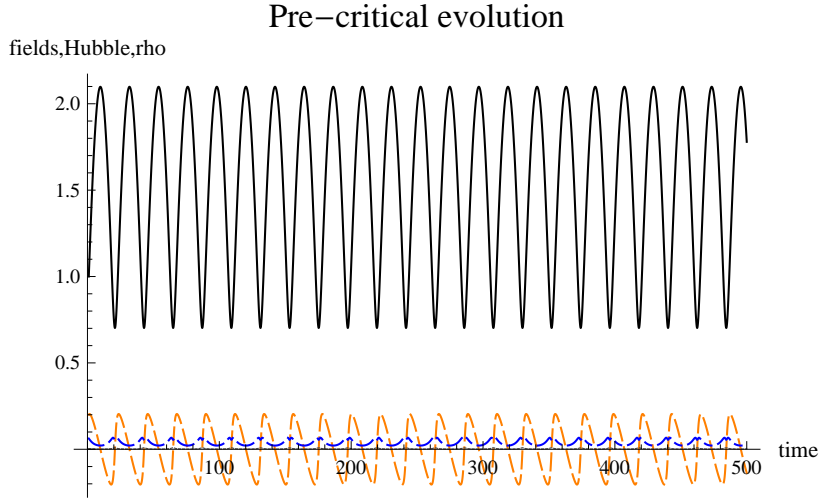}} \hspace*{\fill}\vspace*{1cm}
   \caption{
Solution of exact  coupled field and Einstein equation taking an
effective $\chi$ mass of SSB prior to critical fluctuation:
assumed parameters are $\sqrt{m_{\chi}^2- g \langle \phi^2 \rangle } =  1/5, 
g = 1, \xi_2 =5 , \xi_4=1, \Lambda = 0.1$
in the Planck unit $M_{\rm P} = 1$.
Centrifugal repulsive term in the potential $0.1/\chi^2$ is added.
$\chi$ field in black, its derivative in dashed orange, Hubble rate in dash-dotted
blue, and effective $\Lambda$ in dotted black.
}
   \label {pre-pt evolution}
 \end{center} 
\end{figure*}

Let us work out the critical point of quantum fluctuation for symmetry restoration.
The next leading term,  the term $c_3/(2 \chi^2) $, to the 
leading potential term of $- c_1 \ln \chi$ at the field infinity
(see equations in (\ref{potential term coeffs}), is shifted as
$c_3 \rightarrow \bar{c}_3$;
\begin{eqnarray}
&&
\bar{c}_3 \sim \frac{1}{ 5\xi_4^2} \left( 2 g \xi_2 M_{\rm P}^2 + 3\xi_4 ( m_{\chi}^2
- g \langle \phi^2 \rangle )
\right) 
\,.
\end{eqnarray}
Hence the field can roll down to infinity when quantum fluctuation grows beyond
$ \langle \phi^2 \rangle_{\rm cr}$ given by $\bar{c}_3 =0 $:
\begin{eqnarray}
&&
\langle \phi^2 \rangle_{\rm cr} = \frac{m_{\chi}^2}{g} + \frac{ 2 \xi_2 M_{\rm P}^2}{3 \xi_4}
\,.
\end{eqnarray}
We may call this critical fluctuation.
We illustrate solutions of  exact solutions prior to critical fluctuation
in Fig(\ref{pre-pt evolution}).

As discussed in more detail below, the inflaton mass $M_{\chi}\propto 1/\chi$
defined in (\ref{chi-mass}) varies with
time, ultimately decreasing with redshifts as $\propto (z+1)^2$.
With decreasing $M_{\chi}$, 
the equilibrium potential minimum is then recovered again at $m_{\chi}/\sqrt{g}$,
violating the condition of symmetry restoration 
given by (\ref{symmetry restoration condition}).
This does not pose any problem to the scenario of vanishing cosmological constant just presented,
because once the inflaton field passes over a local potential barrier right to $m_{\chi}/\sqrt{g}$,
the potential monotonically decreases behaving like $- \ln \chi$ towards the field infinity.
What matters is
a temporary change of potential shape around the electroweak
phase transition, and a later shape change in this region is irrelevant to the cosmological
constant problem.

\subsection
{\bf Radiation dominated epoch}

Time evolution in the radiation-dominated universe 
of thermalized particles after inflation
is described by inflaton field equation,
\begin{eqnarray}
&&
\ddot{\chi} + \frac{3}{2t} \dot{\chi} = \frac{g}{5 \xi_4} \frac{M_{\rm P}^4}{\chi}
\,,
\label {rd field eq}
\end{eqnarray}
where the Hubble rate $H = \dot{a}/a = 1/ 2t$, since
the inflaton contribution is negligible at this epoch.
Solutions are searched for by assuming an ansatz, $\chi(t) = A t (\ln t)^p$,
and one arrives at approximate solutions,
\begin{eqnarray}
&&
\chi(t) = \sqrt{\frac{ 2g}{15 \xi_4 }} M_{\rm P}^2 t
\,.
\end{eqnarray}
Namely, the first term $\ddot{\chi}$ is negligible compared to the second
of right-hand side of eq.(\ref{rd field eq}).
Since the redshift factor is defined by $z+1 = a(t_0)/a(t)$ with $t_0$ the present cosmic time,
this gives a simple relation,
\begin{eqnarray}
&&
\frac{\chi(t)}{\chi(t_0)} = \frac{t}{t_0} = (z+1)^{-2}
\,, 
\label{de variation 1}
\\ &&
\chi(t_0) \approx \sqrt{\frac{1 }{ 30}} \sqrt{\frac{g }{ \xi_4}} \frac{M_{\rm P}^2 }{H_0 }
\sim 0.31 \times 10^{60} \sqrt{\frac{g }{ \xi_4}} M_{\rm P}
\,,
\label{de variation 2}
\end{eqnarray}
extrapolating, to the present epoch, the formula $H=1/(2t)$ valid
in the radiation dominated epoch, $t_0 \approx 1/(2H_0)$.
This crude approximation is sufficient for our purpose.

We  summarize the potential of entire scalar system consisting of
O(4) symmetric real inflaton fields having no quantum number
of the electroweak theory  and Higgs doublet.
In the rolling phase towards the field infinity the Einstein frame potential 
of these fields is given by
\begin{eqnarray}
&&
V_{\chi H} = - c_1 M_{\rm P}^4 \ln \frac{\chi}{M_{\rm P}} 
+ \frac{\lambda_H}{4} \, F^{- 2} (| H|^2 - v^2 )^2
\,. 
\label {entire scalar potential}
\end{eqnarray}
In the $\chi \rightarrow \infty$ limit the first term in the right-hand side
dominates over all other terms.

The $F-$factor is a quartic function of $\chi$,
and at the field infinity the factor here is approximately
\begin{eqnarray}
&&
F^{-2} \approx (\xi_4 (\frac{\chi}{M_{\rm P}})^4 )^{ -2 }
\propto (z+1)^{16 }
\,.
\end{eqnarray}
This contribution at earlier epochs is lots larger than at present.
The Higgs boson mass given by (\ref{higgs boson mass vs chi}) thus follows
the rule, $\propto F^{- 1} \propto   (z+1)^{8 }$
in radiation-dominated universe.

\section
{\bf Late time acceleration}

\subsection
{\bf Differential equations and its solutions}

Basic equations have been already derived in 
 (\ref{conformal chi-eq: asymptotic}) and (\ref{einstein eq 1}) with
the effective potential (\ref {effective potential at infty}).
To recapitulate, the equations for spatially homogeneous mode read as
\begin{eqnarray}
&&
\ddot{\chi} + 3 \frac{\dot{a}}{a} \dot{\chi} - 4 \frac{\dot{\chi}^2}{\chi}
=  c_0  \frac{M_{\rm P}^4}{\chi}
\,, \hspace{0.3cm}
c_0 = \frac{g}{5 \xi_4} > 0
\,,
\label {simplified field eq}
\\ &&
(\frac{\dot{a}}{a})^2 = \frac{\rho_{\rm DE}}{6M_{\rm P}^2}
\,, \hspace{0.3cm}
\rho_{\rm DE} \approx
\frac{M_{\rm P}^6}{5 \chi_*^2}
(\frac{ \xi_2}{\xi_4 } +  \frac{3 m_{\chi}^2 }{2 g M_{\rm P}^2 }) 
\,.
\label {simplified einstein eq}
\end{eqnarray}
We took the leading term for $\rho_{\rm DE}$ in the right-hand side of Einstein equation.
The next sub-leading terms shall be discussed shortly.

\begin{figure*}[htbp]
 \begin{center}
 \epsfxsize=0.6\textwidth
 \centerline{\epsfbox{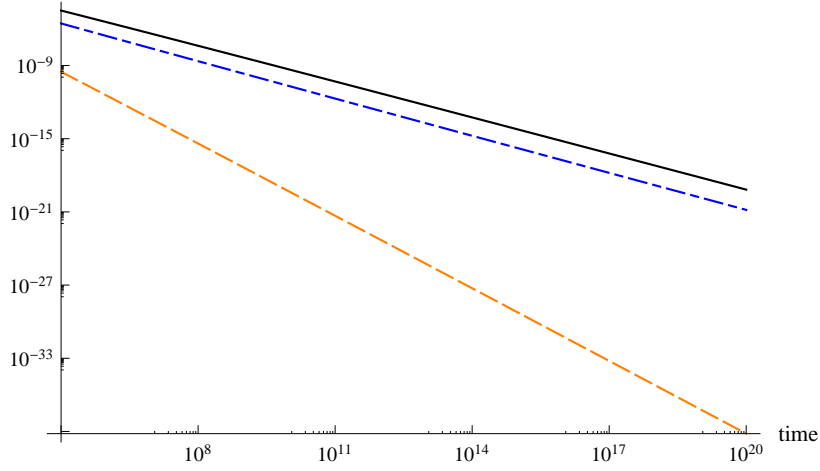}} \hspace*{\fill}\vspace*{1cm}
   \caption{
Potential dominance solution  is used to 
compare three terms in the field equation:
the force term derived from the potential in solid black,
the Hubble term in dashed orange, and the friction term $\propto - \dot{\chi}^2/\chi$
in dash-dotted blue.
Minor parameter dependences are ignored, taking
$c_0  = 1, M_{\rm P} = 1$.
}
   \label {late-time potential dominance}
 \end{center} 
\end{figure*}

There is an important relation that holds irrespective of detailed inflaton evolutionary behavior:
the Einstein equation immediately tells that the universe
is accelerating, and the present value of dark energy density is given by
\begin{eqnarray}
&&
\rho_{\rm DE}(t_0) = 6 H_0^2 M_{\rm P}^2 \sim (2.1\, {\rm meV})^4
\,.
\end{eqnarray}
In the accelerating late-time universe
the inflaton field equation reads as
\begin{eqnarray}
&&
\ddot{\chi} + 3 H_0 \dot{\chi} - 4 \frac{\dot{\chi}^2}{\chi}
=  c_0  \frac{M_{\rm P}^4}{\chi}
\,,
\end{eqnarray}
with a constant Hubble rate $\dot{a}/a \sim H_0$.
One can readily understand the behavior of time evolution
by guessing solutions for the field  and verifying its consistency.
The ansatz of solution is a power law, $\chi = A t^{\alpha}$:
the choice of $\alpha = 1/2$ gives a balance of the Hubble and the potential terms,
to give
\begin{eqnarray}
&&
\chi \approx \sqrt{\frac{2 c_0 }{ 3 H_0}} M_{\rm P}^2 t^{1/2}
\,.
\label {late-time chi}
\end{eqnarray}
Neglected terms are found to be of order $t^{-3/2}$, smaller than
the retained term of order $t^{-1/2}$.
The validity of potential dominance is illustrated in Fig(\ref{late-time potential dominance}).

A more accurate closed form of inflaton field equation is
\begin{eqnarray}
&&
\ddot{\chi} + \frac{3}{\sqrt{6} M_{\rm P}} 
\sqrt{\rho_{\rm DE} + \delta \rho }\, \dot{\chi} - 4 \frac{\dot{\chi}^2}{\chi}
=  c_0  \frac{M_{\rm P}^4}{\chi}
\,,
\\ &&
\delta \rho = \frac{ M_{\rm P}^4}{ \xi_4} 
\left( \frac{5}{2} \frac{\dot{\chi}^2}{\chi^4}
- \frac{g}{5}  \ln \frac{\chi}{\chi_*}
\right)
\,.
\end{eqnarray}
It is   difficult to numerically solve this equation,
but one can conjecture that the solution approaches (\ref{late-time chi}).

\subsection
{\bf Equation-of-state factor}

It may be useful to define the equation-of-state factor $w_{\chi}$ for the inflaton field,
\begin{eqnarray}
&&
w_{\chi}^{(E)} =
\frac{\frac{5}{2 F} ( \dot{\chi}^2-  \frac{(\vec{\nabla} \chi)^2}{3 a^2 } )
- V_{\rm eff}^{(E)}(\chi)}
{\frac{5}{2 F}( \dot{\chi}^2 + \frac{(\vec{\nabla} \chi)^2}{ a^2 } )
 + V_{\rm eff}^{(E)}(\chi) }
\,.
\end{eqnarray}
The energy conservation law for inflaton in conformal gravity is 
\begin{eqnarray}
&&
\dot{\rho_{\chi}} = - (3\frac{\dot{a}}{a} + 4 \frac{\dot{\chi}}{\chi} ) \frac{5}{F} \rho_{\chi}
 (1 + w_{\chi}^{(E)})
\,,
\label {energy conservation law in e-frame}
\end{eqnarray}
replacing $\dot{\rho_{\chi}} = - 3\frac{\dot{a}}{a}\rho (1 + w_{\chi}) $ in general relativity (GR).
This equation was derived by using the field contribution to the energy-momentum
tensor in the Einstein frame,
\begin{eqnarray}
&&
\hspace*{-0.3cm}
T^{(\chi)}_{\mu \nu} = \frac{5}{F} ( \partial_{\mu} \chi \partial_{\nu} \chi 
- \frac{1}{2} g_{\mu \nu} \partial_{\alpha}\chi \partial^{\alpha}\chi )
+ \frac{1}{F^2} V(\chi) g_{\mu \nu}
\,.
\end{eqnarray}

In GR there is a cancellation between a part of kinetic term $\dot{\chi}^2/2$ and 
potential $V $, to give the pressure $ p = w \rho = \frac{(\vec{\nabla} \chi)^2}{6 a^2 }$.
This cancellation does not work in conformal gravity due to the presence of
$F-$factors.
It would be instructive to calculate the energy density and the pressure
for inflaton field in conformal gravity.

The potential was calculated at field infinity and is given in (\ref{effective potential at infty}).
Three terms in the energy density are given by
\begin{eqnarray}
&&
\rho_{\chi} =
\frac{5}{2 \xi_4}\frac{M_{\rm P}^4}{\chi^4}
( \dot{\chi}^2 +  \frac{(\vec{\nabla} \chi)^2}{a^2 })
\nonumber \\ &&
+
\frac{M_{\rm P}^4}{5}
\left( (\frac{ \xi_2}{\xi_4 } +  \frac{3 m_{\chi}^2 }{2 g M_{\rm P}^2 }) 
\frac{M_{\rm P}^2}{\chi_*^2}  - 
\frac{g}{\xi_4} \ln \frac{\chi}{\chi_*}
\right)
\,.
\label {dark energy density}
\end{eqnarray}
With the redshift dependences $\chi  \approx 10^{60} \,M_{\rm P}  (z+1)^{-2}$,
of (\ref{de variation 1}) and (\ref{de variation 2}),
and $\xi_* = O(M_{\rm P})$,
the energy density is dominated by the potential term,
and the pressure term given by the second term is completely
negligible at moderate redshifts.
We thus arrive at the conclusion of effective $w$ factor being $-$1.

Observational data of cosmology is  analyzed using
$\Omega_i = \rho_i 8 \pi G_N/(3 H^2)$
for each component $i$, dark energy, cold dark matter,
baryon, and radiation, that contributes to the energy density
and $w-$factor: $-1, 0, 0, 1/3$, respectively.
In our model the energy and the mass density contain
the unit factor $F^{-1/2}$ for dark energy and dark matter,
and $F^{-\epsilon}$ for baryon and radiation as explained in the next section.
There is no problem to interpret data in terms of $\Lambda$ CDM model
provided a promising candidate of cold dark matter is identified.
The only subtle problem that may arise is derivation of the cosmological bound of neutrino mass,
which should be analyzed with great care.

\section
{\bf Inhomogeneous inflaton modes and candidate cold dark matter}

\begin{figure*}[htbp]
 \begin{center}
 \epsfxsize=0.6\textwidth
 \centerline{\epsfbox{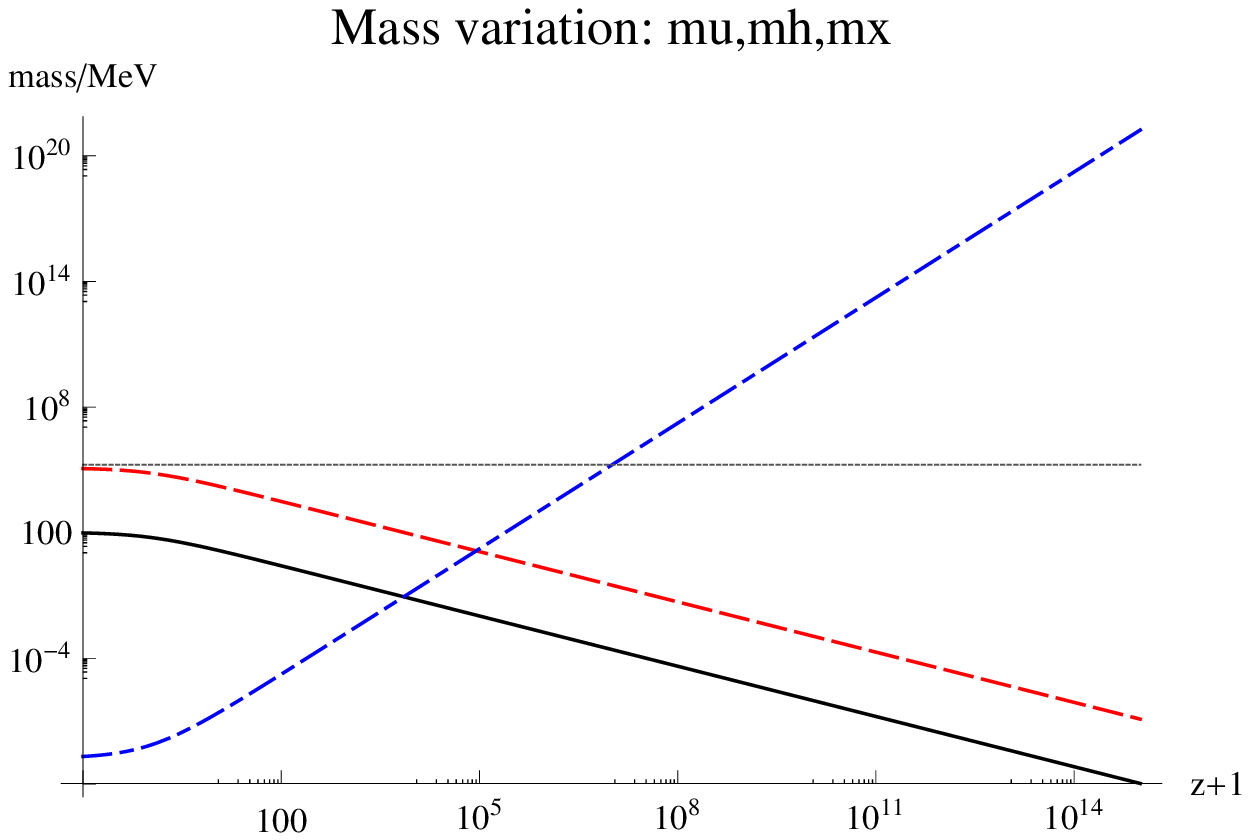}} \hspace*{\fill}\vspace*{1cm}
   \caption{
Mass variation:
muon mass in solid black, Higgs boson mass in dashed orange, and
$\chi$ inflaton mass in dash-dotted blue, plotted against the
redshift factor $z+1$, assuming $\epsilon= -0.1$.
For a reference the line $v= 250 GeV/\sqrt{2}$ is shown in dotted black.
This variation includes the scale change by $1/F$, hence physically
relevant is inflaton mass variation relative to masses of standard model particles.
}
   \label {mass vs redshift}
 \end{center} 
\end{figure*}

\subsection
{\bf Inhomogeneous inflaton modes}

The field equation for spatially
inhomogeneous modes of inflaton can be derived by Fourier-decomposing linearized
general partial differential equation. This becomes possible by exploiting 
 the  space translational invariance under the background field $\chi_b(t)$.
The linearized equation  in terms of deviation $\delta \chi_q
= \chi - \chi_b(t) \propto e^{i\vec{q}\cdot \vec{x}} $ is given by
\begin{eqnarray}
&&
\ddot{\delta{\chi}_q} + 3 \frac{\dot{a}}{a} \dot{\delta{\chi}_q} 
+ \frac{q^2}{a^2} \delta{\chi}_q
= - M_{\chi}^2(z) \delta{\chi}_q
\,, 
\label {inhomo cdm eq}
\\ &&
M_{\chi}(z) = \left(\partial^2_{\chi} V_{\rm eff}^{(E)} \right)_{\chi=\chi_b} = M_{\chi}(0) (z+1)^2
\,,
\\ &&
M_{\chi}(0)  = \sqrt{6} H_0 \sim 2 \times 10^{-33} {\rm eV}
\,, 
\end{eqnarray}
in the radiation-dominated era.

Using $z+1 = \sqrt{t_0/t}$ valid in the radiation-dominated era,
one can analytically solve (\ref{inhomo cdm eq}) in terms of
cylindrical functions $Z_{\nu}(w)$:
\begin{eqnarray}
&&
\delta \chi_q = {\cal A}_q w^{-1/4} Z_{\nu } ( w) 
\,, 
\\ &&
w = 2 q \sqrt{ t_0 t} 
\,, \hspace{0.3cm}
\nu = \frac{\sqrt{ 1 - 16 (M_{\chi}(0) t_0 )^2 }}{2}
\,.
\end{eqnarray}
The value $4 M_{\chi}(0) t_0 = 4 \sqrt{6} H_0 t_0 \sim 100$
gives a purely imaginary $\nu$, the order of cylindrical function.
The solution in the $t \rightarrow 0^+$ limit is $O(t^{- 1/8}) \times$
sinusoidally oscillating function of argument $\Im \nu \ln t/2$.

At large times inflaton amplitudes approach the behavior of
damped oscillation,
\begin{eqnarray}
&&
\hspace*{-0.3cm}
\delta \chi_q(w = 2 q \sqrt{ t_0 t})  \approx  {\cal A}_q \sqrt{\frac{2}{\pi}}
w^{ - 3/4} \cos \left( w - \frac{ 2\nu + 1}{ 4} \pi \right)
\,,
\nonumber \\ &&
\label {cdm sq}
\end{eqnarray}
for $Z_{\nu} = J_{\nu}$, the Bessel function.
A time average of the squared field amplitude over a time span
$T > \pi /(q^2 t_0) $ is
\begin{eqnarray}
&&
\overline{\delta \chi_q^2 } \approx \frac{{\cal A}_q^2}{4 \pi} (2 q \sqrt{t_0 t})^{-3/2}
e^{\pi \Im \nu } \propto t^{-3/4 }
\,.
\end{eqnarray}
This slow decrease behavior differs from the naive expectation 
$\propto 1/a^3 \propto t^{- 3/2}$
for ordinary cold dark matter in radiation-dominated era.
See below on more of this.

In Appendix, we argue that inhomogeneous modes are thermally produced
from ambient particles.
The amplitude ${\cal A}_q$ can then be determined by taking thermal distribution,
\begin{eqnarray}
&&
{\cal A}_q =  \frac{\delta \chi_q(z_{EW})}
{ e^{\sqrt{M_{\chi}^2 + q^2/a^2(t_{EW})}/T_{EW} } - 1}
\,,
\end{eqnarray}
when  inflatons are produced at the electroweak epoch $t= t_{EM}$.
$\delta \chi_q(z_{EW})$ is of order $T_{EW}$.
One can use the small time limit formula of Bessel function traced back from
late times
\begin{eqnarray}
&&
\hspace*{-0.2cm}
\delta \chi_q \approx 2^{-\nu}  {\cal A}_q w^{\nu -1/4} 
\,.
\end{eqnarray}

After their production, decay of inhomogeneous modes is frozen,
and shortly after the electroweak epoch they decouple from
the rest of thermal medium.
The epoch when they become non-relativistic is
estimated by equating the average momentum to the mass ($\sim 2$ meV at
the electroweak epoch), which gives the redshift factor,
\begin{eqnarray}
&&
z_{NR} + 1 = \frac{\pi^4}{ 30 \zeta(3) \sqrt{6}} \frac{T_{EW} }{ H_0 (z_{EW} + 1)^2}
\approx 1\times 10^{14}
\,.
\end{eqnarray}
Thus, it is likely that inhomogeneous inflaton modes become
cold dark matter (CDM) shortly after the electroweak phase transition.

\subsection
{\bf Time variation of CDM energy density}

The formula of inhomogeneous mode energy density is written in terms of field and field derivative;
\begin{eqnarray}
&&
\hspace*{-0.3cm}
\rho_{\rm CDM} = M_{\rm P}^4 \left( \frac{5}{2 \xi_4} \frac{(\dot{\delta \chi})^2}{\chi_b^4}
+ \frac{5}{2 \xi_4} \frac{(\vec{\nabla}\delta \chi)^2}{ a^2 \chi_b^4}
+ \frac{g}{10\xi_4} \frac{(\delta \chi )^2}{\chi_b^2}
\right)
\,.
\nonumber \\ &&
\label {cdm energy density}
\end{eqnarray}
Redshift dependences of the three terms in the right-hand side 
are $(z+1)^{13.5 }\,, (z+1)^{11.5 }\,, (z+1)^{ 5.5} $, respectively.
At  latest epochs the last potential term is dominant.
We can estimate CDM energy density after the epoch of the radiation-matter equality:
\begin{eqnarray}
&&
\rho_{\rm CDM}  (z) \approx \rho_{\rm CDM}  (0)  (z+1)^{5.5}
\,, 
\\ &&
\rho_{\rm CDM}  (0) = \frac{g}{10\xi_4} M_{\rm P}^4\frac{(\delta \chi_0)^2}{\chi_0^2}
\,,
\\ &&
M_{\rm P}^4 \frac{(\delta \chi_0)^2}{\chi_0^2}
\approx \frac{\pi^2 T_{EW}^4}{15} (z_{EW}+1)^{-3/2}
\sim 2 \times 10^{21 } {\rm eV}^4
\,, 
\nonumber \\ &&
\\ &&
\rho_{\rm CDM}  (0) \approx (0.8\, {\rm meV})^4
\,.
\label {present cdm energy density}
\end{eqnarray}
The  estimated present CDM energy  density $\rho_{\rm CDM} (0)  $
falls in an interesting range,  and is
 consistent with the observed dark matter energy density.

A possible problem against identifying inhomogeneous modes as
cold dark matter might be in the decrease rate of the energy density,
$(z+1)^{5.5/2}$, or
$a^{-11/4 }$ in terms of scale factor of radiation dominance.
This is slightly slower than the usual CDM $a^{-3}$ behavior.
This does not seem to be a real problem  in analysis
of observational data in terms of $\Lambda$ CDM model.

There is another candidate of cold dark matter in our model.
It is $\phi-$inflaton.
This field satisfies a non-linear equation of the form,
\begin{eqnarray}
&&
\ddot{\phi} + 3 \frac{\dot{a}}{a}\dot{\phi} 
- \frac{\partial_{\chi} F}{F} \dot{\phi}
- \frac{\vec{\nabla}^2 \phi}{a^2} = \frac{c_-^2}{a^6 \phi^3}- m_{\phi}^2 \phi
\,.
\end{eqnarray}
The potential minimum $\phi_*$ is determined by balancing the centrifugal repulsion
and the ordinary quadratic mass term in the right-hand side.
Around this minimum $ $ given by
\begin{eqnarray}
&&
\phi_* = a^{-3/2} \sqrt{ \frac{c_-  }{ \xi_4 m_{\phi}(z)  } } (\frac{M_{\rm P} } { \chi_b} )^2
\,,
\\ &&
m_{\phi}(z) = m_{\phi}(0) (z+1)^2
\,, 
\\ &&
m_{\phi}(0) \sim \sqrt{30 g} H_0 \sim 5 \times 10^{-33} {\rm eV}
\,,
\end{eqnarray}
one can analyze dumped oscillation in terms of linearized
deviation $\delta \phi = \phi - \phi_*$.
The Fourier mode decomposition 
$\delta \phi = \sum_{\vec{q}} \delta \phi_q e^{i\vec{q}\cdot \vec{x}}$
gives
\begin{eqnarray}
&&
\ddot{\delta \phi_q} + ( 3 \frac{\dot{a}}{a} -  4 \frac{\dot{\chi_b}}{\chi_b}  ) \dot{\delta \phi_q}
+ \frac{q^2}{ a^2} \delta \phi_q = - 4 m_{\phi}^2 \delta \phi_q
\,.
\end{eqnarray}
The $\phi-$field equation is essentially similar to the $\chi-$inflaton
inhomogeneous mode equation, and $\phi$ quanta are thermally produced
in similar fashions to $\chi$ CDM.

Presumably, a more attractive candidate is
gravitationally collapsed clumps made of inhomogeneous modes;
primordial black holes.
As discussed in \cite{next paper} and briefly mentioned in Appendix,
gravity effects measured by $G_N M_{\rm clump}^2 \propto (z+1)^8$ grows
towards early epochs of our universe if the clump mass
$M_{\rm clump}$ is dominated by inhomogeneous inflaton modes.
Primordial black holes of mass $\sim 10^{23}$ gr within the horizon 
at the end of inflation may be copiously produced.
They do not over-close the universe
due to the present upper bound of energy density $\approx (0.8 {\rm meV})^4$
of (\ref{present cdm energy density}).

\section
{\bf Summary}

It is striking that a class of conformal scalar-tensor gravity theories have a rich variety
of physics in the early universe, and has a potential of solving the long-standing problem
of fine-tuned cosmological constant along with a slow-roll inflation,
late-time accelerating universe and cold dark matter.
We shall summarize what we have achieved in our model.

In our class of models there are, in addition to standard model lagrangian functionals,
two polynomial functions, for which we
took quartic polynomials for a simplest successful model.
They are conformal and  potential functions,
and these are functions of four real fields both having O(4) symmetry.
The conformal function $F(\chi)$ is positive definite 
and is a function of inflaton field $\chi$,
coupled to the Ricci scalar of metric field in the lagrangian density.
The potential function has a O(4) symmetric wine bottle shape
and has a minimum away from the field zero origin.
A bare cosmological constant may be present in the Jordan metric frame,
but in the Einstein frame it is multiplied by $F^{-2}(\chi)$, which may be regarded as an effective
cosmological constant variable.

Suppose that the universe started with an inflaton field 
value larger than $O( 5 M_{\rm P})$.
The effective potential derived by integrating the force
term in the inflaton field equation is found to decrease
towards a smaller O(4) symmetry breaking point.
The slow-roll inflation producing a right amount of the spectral index
and satisfying a constraint of the tensor-to-scalar mode ratio 
consistent with observations is realized by choosing parameters 
within a wide range.

Towards the end of inflation the
inflaton field oscillates around the potential minimum. 
Two things happen during this oscillation phase;
firstly, copious particle production including standard
model Higgs boson,
and secondly growing accumulated quantum fluctuation.
These two are related by the fluctuation-dissipation theorem.
Copiously produced particles quickly thermalize among
their own interactions, giving a hot big-bang universe.
Accumulated quantum fluctuation ultimately restores O(4) symmetry,
with inflaton field being pushed back towards the field infinity under a negative
logarithmic potential slope.
Inflaton field rolls ultimately to the infinity, roughly proportional to the  elapsed time,
while the energy density stored in inflaton decreases much
more slowly, behaving effectively as a constant.

Inhomogeneous mode of inflaton field is shown to have a quantum mass
around the  Hubble constant $\sim 10^{-33}$eV  at present, 
and they act as a cold dark matter,
becoming temporarily dominant after radiation dominance.
Finally, the  inflaton field takes over
and late time accelerating universe begins and lasts till the present,
leaving the dark energy density of order, (a few meV)$^4$.

Another attractive candidate of the dark matter is
primordial black hole due to stronger gravity effects
in the early universe.

Baryo-genesis or lepto-genesis can be accommodated 
in a number of ways, with additional physics input beyond the standard model.

Clearly, many details have to be worked out,
but it would be exciting to be able to discuss the entire history
of early universe without worrying about the fine-tuning of cosmological constant.

\section
{\bf 
Appendix: Production rate and decay freeze-out of inhomogeneous inflaton modes}

In order to discuss  production and  a possibility of inflaton decay,
it is necessary to  precisely understand the mass relation of standard model particles,
Higgs boson,  fermions (quarks and leptons), and the inflaton.
A candidate of inflaton cold dark matter (CDM) is identified as the spatially
inhomogeneous mode of inflaton field $\chi$, and its time
evolution is fully discussed  in the text.
We mention here that 
the inflaton mass $M_{\chi}$ is defined by the square root of potential derivative,
$ \partial_{\chi}^2 V_{\rm eff}$ in (\ref{chi-mass}),
and that it  dynamically changes with the cosmic expansion,
being proportional to the square redshift, $ (z+1)^2$.
This time dependence is different from those of standard model particles.
According to the Weyl scaling to the Einstein metric frame,
fermion masses and other boson (Higgs scalar and gauge vector bosons) masses
have different redshift differences, $(z+1)^{16}$ and $ (z+1)^{8}$,
respectively.

But this weird time dependence can be modified and
a new rule may be set up such that all standard model particle masses
are changed according to the same rule, $(z+1)^{8\epsilon}$.
The derivation of modification shall be given separately in
a companion paper \cite{next paper}.
We adopt this modification, since
it may drastically change time evolution of cold dark matter.
At the Feynman rule level, the factor,
\begin{eqnarray}
&&
F^{-2\epsilon}(\chi_b) \approx 
 \left(\xi_4 (\frac{\chi_0}{M_{\rm P}})^4 \right)^{-2\epsilon } (z+1)^{16 \epsilon }
\,,
\end{eqnarray}
is attached to all interaction vertexes.
With the anticipated numerical estimate of $\chi_0 (z+1)^{-2} /M_{\rm P} \approx 10^{30}$,
the $\epsilon$ factor must be negative to avoid a large suppression.
If the inflaton mass is in the denominator of probability amplitude,
a further factor $\propto (z+1)^2$ is present.

We illustrate mass relations assuming a small negative $\epsilon$ value
in Fig(\ref{mass vs redshift}).
From this illustration we arrive at a conclusion that
significant inflaton production occurs via high energy collision of fermions
such as lepton scattering,
$l_1 + l_2 \rightarrow l_3 + l_4 + \chi$ that arises from Higgs $H_0$ exchange
in the t-channel emitting $\chi$ by $\chi H_0 H_0$ vertex in the middle.
The probability amplitude of this process is product of external wave functions
$ \bar{\psi}_3 \psi_1 \, \bar{\psi}_4 \psi_2 \, \varphi_{\chi} $ times
\begin{eqnarray}
&&
\frac{-i \,y_l^2 c_H F^{-6\epsilon}(\chi_b) }
{\left( (p_1- p_3 ) - m_H^2 \right)  \left( ( p_2- p_4) - m_H^2 \right)}
\,,
\end{eqnarray}
where $y_l$ is Yukawa coupling constant to lepton, and
$c_H$ is given in (\ref{chi-coupling}).

We shall be content with an order of magnitudes estimate of production cross section
instead of calculating that  in detail.
From dimensional grounds we expect that the cross section is
of order,
\begin{eqnarray}
&&
\sigma_{\chi} = \frac{y_l^4 c_H^2\, F^{-12 \epsilon}(\chi_b)}{(T^2 + m_H^2)^2}
\,,
\end{eqnarray}
expecting  larger values of the incident energy given by the temperature $T$
and the Higgs boson mass to dominate in the denominator.
Production rate $\Gamma_{ll \rightarrow \chi}$ is given by $\sigma_{\chi} n_{l}$ where
$n_l = 3 \zeta(3) T^3/(2\pi^2 ) \sim 0.18\, T^3$ is the number density of lepton $l$.
The ratio of production rate to the Hubble rate is then estimated as
\begin{eqnarray}
&&
\frac{\Gamma_{ll \rightarrow \chi} }{H} \approx
0.01 (\frac{m_l}{100 {\rm MeV}})^4 (\xi_4)^{ 6 \epsilon+3/2} g^{-12 \epsilon -3/2 }
\nonumber \\ &&
\cdot
\left( 0.31 \times 10^{30} ( \frac{10^{15} }{z+1})^2
\right)^{ - 24 \epsilon -2}
\frac{(T/m_H)^3 }{\left( 1 + (T/m_H)^2 \right)^2 }
\,.
\end{eqnarray}
For instance, the prefactor becomes $\sim 0.8 \times 10^{10}$ if $\epsilon = -1/10$.
Decoupling occurs at a redshift $0.1 \times $ the electroweak value
for this choice of $\epsilon$.
For a larger $-\epsilon$ the decoupling redshift becomes
closer to the electroweak value.
Thus, a range of parameters $\epsilon \leq - 0.1$ gives
production rates strong enough for thermalized inflaton CDM.

On the other hand, the main inflaton  $\chi \rightarrow H_0 H_0$
decay rate compared to the Hubble rate
\begin{eqnarray}
&&
\frac{8 \lambda_H^2}{\pi}  (\frac{v^2}{\chi_0})^2 \frac{1}{H_0}
(\xi_4 \frac{ \chi_0}{M_{\rm P} })^{ - 8\epsilon} (z+1)^{16 \epsilon +2 }
\,,
\end{eqnarray}
is too small with the same choice of $\epsilon = -1/10$ as before
and around the electroweak scale:
the $\chi$ CDM decay is essentially frozen despite that
it is kinematically allowed.

\vspace{0.5cm}
\begin{acknowledgments}
This research was partially
 supported by Grant-in-Aid   21K03575   from the Japanese
 Ministry of Education, Culture, Sports, Science, and Technology.

\end{acknowledgments}


\begin{thebibliography}{99}

\bibitem{lambda-problem}
S. Weinberg,
Rev. Mod. Phys. {\bf 61}, 1 (1989),
and references therein.
There are many other proposals for solving the cosmological constant problem.
Only an incomplete partial list after this review is given in \cite{k-essence}, 
\cite{bousso-polchinski}, \cite{witten}
\cite{cc polyakov}.



\bibitem{k-essence}
C. Armendaritz-Picon, V. Mukhanov, and P.J. Steinhardt,
Phys.Rev.Lett.{\bf 85}, 4438 (2000). 
Phys.Rev.{\bf D63}, 103510(2001); 
{\it Essentials of k-Essence},
arXiv: astro-ph/0006373v1 (2000).

\bibitem{bousso-polchinski}
R. Bousso and J. Polchinski,
JHEP {\bf 6}, 006 (2000);
``Quantization of four-form fluxes and dynamical neutralization of the
cosmological constant'', arXiv: hep-th/0004134 (2000).

\bibitem{witten}
E. Witten: ``The cosmological constant from the viewpoint of string theory'',
hep-ph/0002297 (2000).

\bibitem{cc polyakov}
A.M. Polyakov,
Nucl. Phys. {\bf B797}, 199 (2008).


\bibitem{my21}
M. Yoshimura,
``Bifurcated symmetry breaking in scalar-tensor gravity''
 arXiv: 2112.02835v2 (2021);
Accepted for publication  in PRD.


\bibitem{cosmology}
A standard textbook of modern cosmology is
S. Weinberg, {\it Cosmology},
Oxford University Press, New York (2008).

\bibitem{jbd} 
Historically, the conformal coupling to Ricci scalar was introduced
by the following references, however without the potential term $V(\chi)$.

P. Jordan, Z. Phys. {\bf 157}, 112 (1959).

C. Brans and H. Dicke,
{\sl Phys.\ Rev.\ }{\bf 124}, 925(1961).

\bibitem{kaluza-klein} 
Th. Kaluza, Sitzungber. Preuss. Akad. Wiss. (1920), 966. 

O. Klein, Z. Phys. 37 (1926), 895.

\bibitem{gh unification}
N. Manton, Nucl. Phys. B158 (1979) 141.

D.B. Fairlie, Phys. Lett. B82 (1979) 97; J. Phys. G5 (1979) L55.

P. Forgacs and N. Manton, Comm. Math. Phys. 72 (1980) 15.




\bibitem{nucleo-synthesis bound}
R.A. Malaneya and G. J. Mathews,
Physics Reports {\bf 229}, 145 (1993);
``Probing the early universe: a review of primordial nucleosynthesis 
beyond the standard big bang''


\bibitem{gr tests}
C.M. Will,
 Pramana {\bf 63},  731  (2004);
{\it The Confrontation between General Relativity
and Experiment},
arXiv:gr-qc/0103036v1 (2002).




\bibitem{inflation models 1}
A. Linde,
Phys. Lett. {\bf B108}, 389 (1982);
{\bf B114}, 431 (1982).
A. Albrecht and P. Steinhardt,
Phys.Rev.Lett. {\bf 48}, 1220(1982).

\bibitem{kami-kov}
M. Kamionkowski and E.D. Kovetz,
Annual Review of Astronomy and Astrophysics,
{\bf 54}, 227 (2016), arXiv:1510.06042.

\bibitem{planck}
Planck Collaboration 2018 \lromn6,
Astron. Astrophys. {\bf 641}, A6 (2020);
arXiv:1807.06209.


\bibitem{bicep/keck}
BICEP/Keck Collaboration,
P.A.R. Ade et al,
Phys. Rev. Lett. {\bf 127}, 151301 (2021);
arXiv:2110.00483 (2021).



\bibitem{preheating 2}
L. Kofman, A. Linde, and A.A. Starobinsky, 
{\sl Phys.\ Rev.\ Lett.\ }{\bf 73}, 3195(1994).
Phys. Rev. {\bf D 56}, 3258(1997).

\bibitem{my95}
M. Yoshimura,
Progr. Theor. Phys. {\bf 94}, 873 (1995).
This paper discusses explosive particle production
caused by external oscillators like periodic inflaton oscillation, 
leading to parametric amplification, and was applied to the reheating problem
after inflation.
The formalism developed there may actually
be applied to oscillators of general time dependent frequency
such as the present model.

H. Fujisaki, K. Kumekawa, M. Yamaguchi, and M. Yoshimura,
Phys. Rev. {\bf D53}, 6805 (1996).

M. Yoshimura, hep-ph/9605246.


\bibitem{next paper}
M. Yoshimura,
``Stronger gravity in the early universe'',
to be updated to arXiv as a companion paper with the present work.
A resolution is proposed there to modify the mass relation in conformal gravity
theories by a consistent formulation  based on the first principles of quantum
field theory.
Resulting consequences show that 
mass ratios among standard model particles do not evolve with time,
while the mass ratio of inflaton to standard model particles necessarily evolve.






\end{thebibliography}
\end{document}